\documentclass{article}

\newcommand{\ie}{\mbox{i.\,e.\,\ }}
\newcommand{\iec}{\mbox{i.\,e.\,}}
\newcommand{\role}{r\^{o}le }
\newcommand{\eg}{\mbox{e.\,g.\,\ }}
\newcommand{\egc}{\mbox{e.\,g.\,}}
\newcommand{\etc}{etc.\,\ }
\newcommand{\tnsr}[1]{\ensuremath{\mathbf{ #1 }}}
\newcommand{\mc}[1]{\ensuremath{\mathcal{#1}}}
\newcommand{\ket}[1]{\ensuremath{\left|  #1 \right\rangle}}
\newcommand{\hilbert}[1]{\ensuremath{\mathcal{#1}}} 
\newcommand{\op}[1]{\ensuremath{\widehat{\textsf{\ensuremath{#1}}}}}
\newcommand{\id}{\op{\mathsf{1}}}
\newcommand{\denop}{\ensuremath{\rho}}
\newcommand{\tr}{\textsf{Tr}}

\newcommand{\twk}[2]{\ket{\begin{array}{c}\mbox{#1} \\ \mbox{#2}
\end{array}}}

\usepackage{array}
\usepackage{relsize}
\usepackage{chicago}
\usepackage{supertabular}

\makeatletter
\renewcommand{\section}{\@startsection
   {section}%
   {1}%
   {0mm}%
   {-\baselineskip}%
   {0.5\baselineskip}%
   {\bfseries\normalsize\centering}}%
\makeatother

\begin{document}

\begin{center}
\LARGE
\textbf{Worlds in the Everett interpretation}

\vspace{0.3cm}

\textbf{\textit{David Wallace$^*$}}

\begin{figure*}[b]
(\textit{March 15, 2001})

* Centre for Quantum Computation, The Clarendon Laboratory, 
University of Oxford, Parks Road, Oxford OX1 3PU, U.K.

(\textit{e-mail:} david.wallace@merton.ox.ac.uk).

\end{figure*}

\normalsize
\end{center}

\vspace{0.7cm}

\begin{quote}
This is a discussion of how we can understand the world-view given to us by the 
Everett interpretation of quantum mechanics, and in particular the \role 
played by the concept of `world'.  The view presented is that we are entitled 
to use `many-worlds' terminology even if the theory does not specify the worlds 
in the formalism; this is defended by means of an extensive analogy with the 
concept of an `instant' or moment of time in relativity, with the lack of a 
preferred foliation of spacetime being compared with the lack of a preferred basis 
in quantum theory.  Implications for identity of worlds over time, and for 
relativistic quantum mechanics, are discussed.

\end{quote}

\vspace{0.4cm}

\section{Introduction}

If we take unitary quantum mechanics seriously \textit{\'{a} la} 
Everett~\citeyear{everett}, the view of reality which emerges is certainly not 
that of a single classical universe.  But is it a `many-worlds' view (as 
\citeN{dewitt} and \citeN{deutsch85} have argued)?  We can object to this view 
\emph{taken literally} on many grounds: ontological extravagance, difficulties 
with relativistic covariance, violation of the principle of the identity of 
indiscernibles, the need to find an exactly preferred basis, etc.\footnote{See \citeN{barrett} for details of these criticisms.}  But if we reject 
such a view then how are we to understand the Everett interpretation --- what 
metaphysical picture does it give us?  

Looked at another way, defenders of an Everettian viewpoint face a dilemma: just 
how seriously are we to take these worlds?  On the one hand, if we were to take them 
literally and build them into our formalism then we would face the preferred-basis 
problem at its worst.  Decoherence would be no use to us, for we would need an \emph{exact} 
world-defining principle and not some pragmatic criterion.  Furthermore,
there seems to be no relativistically covariant way to define a world, so this approach
would push us towards the same problems with relativity faced by approaches like the 
pilot-wave theory \cite{bohm,holland} or state-reduction theories
\cite{GRW,pearle}.  On the other hand, if we banish these worlds from our 
formalism we must answer the criticism that our theory is just uninterpreted 
mathematics, and explain how sense can be made of the universal state.\footnote{In 
either case, of course, we must explain how to understand probability;
but that problem lies beyond the scope of this paper.}

In this paper I shall attempt to defend three claims.  The first is that
Everettians can avoid this dilemma by a compromise: they may
legitimately and meaningfully use the terminology of many worlds without
being required to represent these worlds in their formalism.  The second
claim is that this view of Everett, rather than being either
contradictory or a pathology confined to quantum mechanics, is typical
of what we should expect when interpreting a theory which changes our
concept of what the world is.  The third claim is that developing such a
view of quantum mechanics brings out a number of deep analogies with the
other twentieth century physical theory which radically changed our
concept of the world: relativity. 

The second of these claims is defended in section \ref{theories}, in which I shall argue that any theory 
which revises our previous ways of understanding reality 
nevertheless needs to make contact with these previous theories if we are to 
be able to comprehend the new theory.  These previous ways are wrong in some 
respects (else we would not need the new theory!) so their use in an interpretation 
of the new theory may be incomplete or obscure; nonetheless such use is essential 
if the new theory is to be seen as a theory of \emph{physics}.  We may hope 
eventually to come to understand a given theory without describing any of its 
concepts in this way, but we cannot begin without any reference to what has gone 
before.

In section \ref{spacetime} the classical concept of relativistic 
spacetime is used as a concrete example of this process, but the main reason for discussing spacetime 
extensively is to set the stage for a discussion of the profound analogies between (relativistic) spacetime 
and the 
Everettian universal state.  These analogies have already been discussed --- 
notably, and extensively, by 
Simon Saunders~\citeyear{saundersevolution,saundersdecoherence,saunderstense,saundersmetaphysics,saundersprobability} --- in 
the context of the probability problem in Everett interpretations.\footnote{The problem 
here is (at least in part) how to do without a criterion by which we can
individuate worlds and identify how they change over time, or how to
provide one if it turns out to be indispensable.  Saunders, and
Tappenden~\citeyear{tappenden2000,tappenden1999}, have argued that such a criterion 
is not needed, and this view will be adopted here (see however
\citeN{barrett} for a dissenting view, and \citeN{butterfield1996} for
further discussion).}  My approach is 
complementary to this (and I mention probability only in passing): if we accept the 
Everett interpretation, we may use the spacetime analogy to cast light on how we 
can describe and understand the world-view it gives us.  (A sort of converse also holds:
if we insist that
the Everett interpretation is meaningful only with a preferred basis
then we are forced to accept that relativistic spacetime is meaningful
only with a preferred foliation.)

In the remainder of the paper I shall attempt to sketch out this approach, and in doing so to defend my first and third claims.  In section
\ref{qm} I address the problem of how we make sense of the Everett
interpretation without an exactly preferred basis, while in section
\ref{persist} I discuss the question of the persistence of worlds
through time and in section \ref{observe} I discuss the role of the observer
in the Everett interpretation, and contrast the 
approach of this paper with many-minds approaches.  In sections 
\ref{covqm}--\ref{dyn} I consider how 
we are to understand (special-)relativistic quantum theory, and 
in sections \ref{table}
and \ref{concl} I summarise the approach.

\section{Interpreting our Theories}\label{theories}

What is spacetime?  A mathematically motivated answer (see, e.\,g.\,, \citeN{wald}) 
might go something like
\begin{quote}
Spacetime is (or possibly, `is isomorphic to') a set 
$\{\mc{M},\tnsr{g},\tnsr{\phi}_1, \ldots \tnsr{\phi}_n\}$, where $\mc{M}$ 
is a connected smooth paracompact 4-manifold, $\tnsr{g}$ is a smooth symmetric 
2-tensor field on $T\mc{M}$, and the $\tnsr{\phi}_i$ are further tensor fields 
on $\mc{M}$.

\end{quote}

But \emph{given} such a mathematical description, how are we to understand what 
spacetime is?  Assuredly we have to get such an understanding, for without any way 
to interpret the mathematics as a description of the world our theory can have 
neither predictive nor explanatory power.  After all, suppose that we had instead 
said that
\begin{quote}
Spacetime is a set containing exactly six elements, together with a commutative 
group operation on that set of elements.
\end{quote}
To object that this is nothing like spacetime, we need some inkling of what sort of 
object this `spacetime' actually is.

This sort of requirement is general to mathematical physics.  To interpret a 
mathematically-formulated theory as physics we need some way of linking the 
mathematics to the physical world.  At the least, this requires that we locate 
\emph{ourselves} somewhere in the theory, or else that we have some
idea of how the objects and concepts of our everyday world are mapped
into the mathematics.
We are able to accept that, in the light of new 
theories, these objects and concepts may turn out to be not exactly what they seem 
--- and indeed some may be completely illusory --- but our theory needs at least to 
make contact with them in order for us to do physics.

It is, however, important to remember that `our everyday world' is not something to 
which we have direct sensory access.  The existence of a three-dimensional spatial 
universe in which exist various macroscopic objects --- including ourselves --- and 
which we learn about through vision, touch \etc is, rather, an extremely effective 
theory (the `everyday theory'), which explains and unifies our observations.  This 
point is
familiar from Quine's writings (see, \egc, \citeNP{quine}).

At first sight, it then seems that to interpret theories of physics by finding the 
everyday world within them is 
to court infinite regress, for we would seem required in turn to interpret the 
`everyday theory' in terms of something yet more basic. 
(After all, once we have accepted it \emph{as} a theory we 
may formalise it in mathematical language every bit as forbidding as that used to 
describe spacetime.\footnote{``Everyday space is a smooth three-manifold together 
with a smooth symmetric two-tensor $\tnsr{g}$, whose unique associated 
metric-compatible symmetric connection has zero holonomy, \ldots''.}  As such, 
if we are required to explicate spacetime then why not our theory of space?)
It might appear that we would instead do better to link our new theory directly to 
our observations --- presumably if the new theory is an improvement upon the 
`everyday theory' it too will explain our observations, only better.

The problem with this idea is that our observations are themselves inextricably 
entwined with the theory of the everyday world.   To take vision as an 
example,\footnote{This discussion is based on material 
taken from \citeN{dennettconsciousness}.} it is a gross over-simplification to 
suppose that our visual field is simply a region of coloured patches rather like a 
television screen. Much low-level processing of visual data takes place before it 
is presented to higher-order subsystems of the brain and begins to affect our 
actions (we might say `before we are conscious of it').  For a start we have two 
eyes, and the parallax calculations that tell us the distance to nearby objects 
appear to be made at a fairly early stage of processing.  We do not consciously 
work out distances to objects, we are simply aware of their distances in addition 
to their angular positions.  But there is far more processing than this: 
identification of objects, making assumptions about regions of the visual field 
that cannot be observed (such as the blind spot) etc.\,,\ so the 
information presented is 
probably better described along the lines of ``there's a book-case 
at 3 o'clock, 4 metres away to judge by the parallax, five shelves, lots of 
multi-coloured books, all about thirty centimetres high'' than by a pixel-by-pixel 
description of the bookcase and the rest of the visual field.

So there are no theory-neutral observations: rather, there is an existing 
theory in terms of which our observations are automatically interpreted, and which 
we must take as our starting point when interpreting the theories of physics.\footnote{This is 
not a novel view; for an example of a similar viewpoint, consider Einstein (reported
in \citeN[pages 63--64]{heisenberg}):
\begin{quote}
It is quite wrong to try founding a theory on observable magnitudes
alone.  In reality the very opposite happens.  It is the theory which
decides what we can observe.  You must appreciate that observation is
a very complicated process.  The phenomenon under observation produces
certain events in our measuring apparatus.  As a result, further
processes take place in the apparatus, which eventually and by
complicated paths produce sense impressions and help us to fix the
effects in our consciousness.  Along this whole path---from the
phenomenon to its fixation in our consciousness---we must be able to
tell how nature functions, must know the natural laws at least in
practical terms, before we can claim to have observed anything at
all.  Only theory, that is, knowledge of natural laws, enables us to
deduce the underlying phenomena from our sense impressions.  When we
claim that we can observe something new, we ought really to be saying
that, although we are about to formulate new natural laws that do not
agree with the old ones, we nevertheless assume that the existing
laws---covering the whole path from the phenomenon to our
consciousness---function in such a way that we can rely upon them and
hence speak of ``observation.'' 
\end{quote} }
It should come as no surprise that we are capable of comprehending \emph{this} theory 
directly, for  we are evolved to live in a universe described reasonably well by 
it, and so it is hardwired into our brains.  New theories of physics do not have 
this advantage: perhaps through long exposure to them we might come to have a 
similarly direct grasp of them (rather than understand them only via the existing 
theory) but this may not be possible at all and certainly will not be possible when 
we are first presented with the new theories.

To summarise, then, our program for interpreting a 
new theory is inevitably dependent on an existing theory whose interpretation we 
are assumed to understand already (such an existing theory will either be our 
everyday theory of bodies, or will in turn be interpreted by means of that theory).
The requirement is that certain parts of the 
new theory must be `approximately isomorphic' to the existing theory.\footnote{The 
reader may, if desired, interpret this as saying that models of the existing theory 
must be approximately isomorphic to parts of models of the new theory.}  The
isomorphism may be imperfect, revealing parts of our old theory to be in error, and 
our understanding of the old theory may be greatly transformed, but it seems that a 
recognisable shadow of the old theory must exist in the new one for it to be more 
than just mathematics.

(For reasons of space I shall not develop the links between this material and the 
closely related questions of how we are to understand theory change and in what 
sense one theory can be contained within another.   The approach adopted here, 
however,  has much in common with the structural realist program in the philosophy 
of science (see \citeNP{worrall,psillos,ladyman}).)

\section{Relativistic Spacetime and the `Many-instants' 
Interpretation}\label{spacetime}

To illustrate this view of theories, let us return to the question raised at the 
start of the last section: how are we to understand (Minkowski) spacetime?\footnote
{Throughout this paper, `spacetime' means Minkowski 
spacetime, though most of what I have to say applies to any relativistic spacetime.
When quantum field theory is discussed, however, the spacetime metric will be assumed
to be fixed and classical: quantum gravity will not be 
discussed.}  One natural way to do so makes essential use of the concepts of space 
\emph{and} time, regarded initially as separate.  One can say of any given event 
that it is in a given place, at a given time --- the former property being 
represented by a point in three-dimensional space, the latter by a single number.  
As a matter of mathematics, then, we can construct a four-dimensional `space' 
(using the word in its mathematical sense) which consists of  all points in space, 
at all times.  This space is a stack of the various spaces each of which is space 
as we ordinarily understand it, at a given time.   This might be called a 
`many-instants' interpretation of spacetime: spacetime is the collection of all 
instants of time, together with a metric structure connecting them and representing 
the time elapsed between instants.  

On understanding better this metric structure --- specifically, on recognizing its
Lorentzian symmetries --- we see that 
this description of spacetime obscures its nature in many 
ways. For a start, it is clear that any given way of slicing it up is arbitrary, 
and that many other choices of simultaneity could be made. (In more mathematical 
language 
we are saying that given a relativistic spacetime, we can make sense of what that 
spacetime is by considering a global foliation --- but that no such foliation is 
the preferred choice.) 
This becomes worse when we start to interpret simultaneity in the spirit of general 
relativity: that is, entirely as an artificial construction.  We then realise that 
there are a multitude of spacelike slices, identical in our vicinity, which differ 
from one another at greater distances.

The situation seems to worsen when we remember the theoretical nature of our 
concept of space.  After all, the surface we perceive through our eyes is not 
spacelike at all: it is the past light-cone, which does not bear much conceptual 
resemblance to our ordinary concept of space (in particular, it has wholly the 
wrong causal structure).  And furthermore, our everyday understanding of `space' 
comes with the understanding that we are represented in it, that our conscious 
thoughts supervene on some region of space.  But barring radical dualism, our 
thoughts presumably must supervene on some processes in our brains --- and 
processes take finite time to occur, and occupy finite space.\footnote{I regard
this point as uncontroversial.  I do  
not claim that this finite extent of (the object of supervenience of) 
our thoughts is of the order of the `specious present'  --- it may well be that
the `finite time' of these processes to be measured in nanoseconds --- 
but if we want to hold to any sort of functionalist view of the brain then 
we're going to be stuck with \emph{some} finite duration.  

I suppose it's possible, given our limited understanding of the nature of 
consciousness, to imagine that our thoughts actually supervene on instantaneous 
states of the brain, rather than processes.  But this seems highly implausible: 
the brain still has nonzero spatial extent so the objects of supervenience would 
have to be some specific 3-slices of the world-tube of the brain, and it is hard 
to imagine a plausible method to describe \emph{exactly} which slicing to choose 
--- eventually any method based on, say, Lorentz frames will need to make arbitrary 
choices of exactly how we define the brain, and in any case it seems hard to 
motivate such an exact rule.  It might be more plausible for the objects of 
supervenience to be only approximately defined, in which case they could perfectly 
well be states of the brain --- after all neurologists talk perfectly sensibly 
about `the current state of a brain' without worrying about the difficulties of 
defining \emph{precisely} the choice of simultaneity hypersurface which they wish 
to use --- but in a certain sense we'd still supervene on many instants at once.  If 
we do wish to defend a claim of supervenience on an instant, a substantial 
metaphysical package would seem to be required --- such as that developed by 
\citeN{barbour99}.} 
Hence our thoughts supervene on regions of spacetime --- so if we regard 
spacetime as a collection of instants, we have to face up to the fact that a single 
conscious thought supervenes on many instants (`many moments of time') rather than 
just one.

So, as a description of spacetime the many-instants interpretation leaves much to be 
desired: it contains arbitrariness, obscures structural features, and we ourselves 
cannot be said to exist in any single given instant.  Nonetheless it is a description 
with merits.  Specifically, it is a \emph{complete} description: once we have 
specified all the contents of all the instants of time and the temporal relations 
between instants, we have the entire spacetime.  Further, we have an existing 
intuitive grasp of what an instant of time is like, and from this we can gain some 
understanding of what sort of entity spacetime is.  This understanding is enough to 
make contact between our experiences and the formalism of relativity, granting the 
theory predictive and explanatory power.

Furthermore, for everyday purposes (\ie when we confine ourselves to regions of 
modest size and modest velocities) there is a pragmatically preferred foliation, a 
pragmatically preferred way to regard spacetime as a collection of instants: that 
foliation defined by the Lorentz frame appropriate to our current state of motion.  
Granted that it is irrelevant whether at large distances we choose to let the 
foliation deviate from the Lorentzian choice; granted that if we look closely 
enough we will need to make arbitrary choices as to whether to take a Lorentz frame 
appropriate to, say, the velocity of Oxford or of a plane flying over it.  For 
everyday purposes the preferred foliation is effectively well-defined.   It is 
here, in the elements of this foliation, that we find those entities approximately 
isomorphic to the three-dimensional spaces of our everyday world-view.

Note here how important it is that we identify our everyday view of the world as a 
theory: if we did see ourselves as having direct access to it we would have to map 
it onto the past light-cone and not some spatial hypersurface.  As it is, though, 
we can claim to have identified ordinary space (approximately) in our new theory, 
which tells us how in the new theory to describe the universe we observe and how 
to locate ourselves in the theory.  

To say that spacetime
\begin{quote}
is an entity with such-and-such mathematical description \textit{which can be 
thought of as a collection of all instantaneous spaces together with the metric 
relations between them, no matter that it may be decomposed into such a collection 
in many different ways}
\end{quote}
is an inelegant way of describing it.  But it is an accurate one, and since Nature 
and our early education have not seen fit to provide us with a direct intuitive 
grasp of what spacetime is we must begin with such inelegant ways to understand it.
Ideally we might, through working with the spacetime concept, develop a direct 
understanding of it, in which case this ladder can be kicked away.  But I 
doubt that we are yet sufficiently at ease with the concept to allow this.

\section{Quantum Theory}\label{qm}

The connection with Everett should be fairly clear.  Everett's essential postulate 
is that the quantum state, and its unitary evolution, are universal; but this leaves 
us with another forbiddingly unintuitive mathematical description~\cite{haag}:
\begin{quote}
The universal state is a set $\{\hilbert{H},\mc{M},\op{\phi}_1, \ldots 
\op{\phi}_n,\denop\}$ where $\hilbert{H}$ is a separable Hilbert space, 
$\mc{M}$ is a spacetime, the $\op{\phi}_i$ are (distributional) maps from $\mc{M}$ 
to the algebra of bounded operators on $\hilbert{H}$, and \denop\ is a self-adjoint 
bounded operator on $\hilbert{H}$ of trace 1.
\end{quote}
Without some idea of how to interpret this entity we clearly cannot treat it as 
physics.

In trying to understand what the Everett interpretation implies we typically begin 
in the middle, presuming various concepts already understood.  We might for 
instance begin with Schr\"{o}dinger's cat, and suppose some subsystem of the universe
to be described by a Hilbert space $\hilbert{H}$ which is the tensor product of 
spaces describing a radioactive atom, a Geiger counter, a cat, and some auxiliary 
apparatus (the box, the cat-killing device, etc.).  We also suppose that there are 
some states in these component spaces which represent definitely decayed or 
undecayed atoms, triggered or untriggered counters, dead or alive cats, and so 
forth.  

Note how extensively we use the concepts of our previous theory.  In particular we 
suppose that we know what is meant by an alive or dead cat, or a triggered or 
untriggered counter --- it is superpositions of systems in such states that we want 
to understand.

Following the steps of Schr\"{o}dinger's thought experiment, we know that the 
unitary dynamics will lead to some evolution like\footnote{The reason for distinguishing
``counter untriggered'' from ``counter still not triggered'' (and
similarly for the cat and the vial) is somewhat pedantic: being
macroscopic systems, these objects will inevitably evolve to some
degree over any time interval (the cat will metabolise food, for
instance).  In fact, strictly speaking it is unlikely to be correct even
to regard these macroscopic objects as pure states: realistically they
will be significantly entangled both with the environment and with each
other.}
\begin{eqnarray}\label{cat}
\lefteqn{
\twk{atom}{undecayed}
\! \twk{counter}{untriggered}
\! \twk{poison}{in vial}
\! \twk{cat}{alive}
\longrightarrow
}
\\
\nonumber
\\
\nonumber
\frac{1}{\sqrt{2}} 
\lefteqn{
\left(
\twk{atom}{decayed}
\! \twk{counter}{triggered}
\! \twk{poison}{released}
\! \twk{cat}{dead}
+
\right.
}
\\ 
& &
\left.
\twk{atom}{undecayed}
\! \twk{counter still}{not triggered}
\! \twk{poison still}{in vial}
\! \twk{cat still}{alive}
\right) \nonumber 
\end{eqnarray}

Further, our prequantum theories give us adequate understanding of the processes
\begin{eqnarray*}
\lefteqn{
\twk{atom}{undecayed}
\! \twk{counter}{untriggered}
\! \twk{poison}{in vial}
\! \twk{cat}{alive}
\longrightarrow
}
\\
\nonumber
\\ 
& &
\twk{atom}{undecayed}
\! \twk{counter still}{not triggered}
\! \twk{poison still}{in vial}
\! \twk{cat still}{alive}
\end{eqnarray*}
and

\begin{eqnarray*}
\lefteqn{
\twk{atom}{decayed}
\! \twk{counter}{untriggered}
\! \twk{poison}{in vial}
\! \twk{cat}{alive}
\longrightarrow
}
\\
\nonumber
\\ 
& &
\twk{atom}{decayed}
\! \twk{counter}{triggered}
\! \twk{poison}{released}
\! \twk{cat}{dead}.
\end{eqnarray*}
It is then tempting to interpret (\ref{cat}) as describing the evolution of two 
(or two sets of identical) parallel worlds, both primarily evolving in a 
classical way but sometimes interfering with one another. This temptation is 
strengthened when we put ourselves into the formalism: a human observer of the 
cat becomes entangled with it, and the universal state is a superposition of 
definite observations.

As was mentioned in the Introduction, there are serious problems with taking this 
parallel-worlds viewpoint literally; let us see how these play out in
more detail.  The greatest difficulty is probably the need to specify which states 
describe single worlds and which describe superpositions of them.  The only 
way to answer this question --- if we take the individual worlds as ontologically 
primary --- is to specify a `preferred basis' ($\{\ket{i}\}$, say) for the 
universal Hilbert space.  Then any state \ket{\psi} can be expressed in terms of 
this basis:
\[
\ket{\psi} = \sum_i \alpha_i \ket{i}
\]
and interpreted as a collection of parallel worlds (indexed by $i$) each of 
weight $\alpha_i$. 

Four problems present themselves.  Firstly what do we mean by the 
phrase ``each of weight $\alpha_i$''?  It is not sufficient to take it as saying 
that the fraction of world $i$ in the collective is $|\alpha_i|^2$, since this 
discards the relative phases of the component worlds, which play an important 
\role\ in the theory.  

Secondly, how is this $\{\ket{i}\}$ basis selected?  To 
justify the reasoning which led us to Everett in the first place, it had better 
be a basis whose members are not states describing macroscopic objects as 
delocalised, but this restriction is far from enough to specify it.  Making some 
choice by fiat is hard to motivate: the configuration-space basis is one obvious 
choice, but violates relativistic covariance by requiring a preferred reference 
frame.\footnote{There's also a slightly subtler problem with this choice: 
generalised to field theory, the ``position'' basis becomes the basis of 
definite field configurations.  But this basis fails to describe macroscopic 
objects properly, since they are made up of \emph{particles}, which are specific 
superpositions of field-configuration eigenstates.  Choosing a basis of definite 
particle locations in QFT is also problematic~\cite{simonloc}, because (i) the 
concept of location is only approximate for relativistic particles; (ii) given 
mass renormalisation, which states describe particles depends on the energy at 
which we observe; (iii) in general spacetimes, the particle basis cannot be 
exactly defined even for free quantum fields.}

The third and fourth problems concern the fact that these `worlds' do not have all 
of the properties of worlds in the 
prequantum sense.  Our third problem is that they are instantaneous: we may 
decompose the universal state into worlds at some given instant of time, but we 
cannot track the individual worlds when we evolve the state forward in time 
in any satisfactory way.  All we can say is that world 14 increases in weight, 
while worlds 15 and 16 decrease and world  17 changes phase, etc. but we
have no classical notion of one set of definite properties passing into another.
(To see this, suppose that the state \ket{\psi}, above, has evolved over
time $t$ into a new state
\[\ket{\psi'} = \sum_i \alpha'_i \ket{i};\]
if each \ket{i} represents a world which evolves over time then we need some
notion that world \ket{i} has evolved over time $t$ into some other world \ket{i'},
but the formalism of quantum mechanics will not give us this --- it just 
tracks the change in weights of the time-invariant states \ket{i}. To recover such a 
notion would (as \citeN{barrett} has argued) require us to 
supplement unitary quantum mechanics with not only a preferred basis, but also 
some auxiliary dynamical rules --- but by this stage we are far from the Everett 
interpretation \textit{per se}.)

The fourth problem is that our thoughts do not supervene on individual worlds 
any more than they do on individual moments of time: whether we identify the 
objects of supervenience as states or processes of the brain, the projector 
onto (neurologically) functionally identical brain states will be very 
high-dimensional --- for how can it matter to neurology whether a certain electron 
in my frontal lobe is displaced by $10^{-15}$ metres?  For all the richness of 
human experience, there surely cannot be enough distinct conscious states to correspond 
one-to-one with the number of orthogonal states corresponding to a working brain;
I shall argue that (at least if we are functionalists) this forces us to 
identify thoughts as supervening 
on facts about certain regions in large numbers of similar but 
non-identical worlds.\footnote{Such an identification has previously been argued for 
by \citeN{donald}:
\begin{quote}In as far as there are such sets of \emph{identical} experiences, I 
would associate them with the \emph{same} mind. [Donald's emphasis]
\end{quote}}

Why are we forced to such an identification?  If the worlds did not interact at all then 
we could possibly avoid it: it is at least defensible to suppose that there exist 
many distinct non-interacting worlds containing many distinct but functionally 
identical copies 
of me.\footnote{In fact, this viewpoint is held by modal realists such as 
\citeN{lewisplurality}.}  The problem is that for any simple choice of the 
preferred basis, interference between different terms in the basis is vital to the 
dynamics of the processes in my brain (or indeed any macroscopic object).  
If we choose the position basis as 
preferred, for instance, we might decompose a certain brain state (describing my 
brain in a certain definite configuration) into position eigenstates and suppose 
that each is a separate conscious entity.  But if one such eigenstate were isolated 
from the rest (if an external observer measured the position of all brain 
constituents, for instance) then that state would evolve rapidly into organic soup 
and certainly would cease to be a functioning brain.\footnote{In case this is not 
obvious, consider the bonds holding together the molecules of the brain: each is 
made up of various electrons in highly delocalised states.  A localised electron, 
then, is conversely in a superposition of binding, and not binding, a given pair 
of atoms.  Breaking a significant fraction of the bonds in every complex molecule 
in the brain would probably not be the only deleterious effect of a position 
measurement, but it will do for a start.}  Hence if mental facts are supervenient 
on facts about brain configurations in 
single worlds then each of us remains a thinking 
being only because of constant interference from the particles comprising the 
brains of countless neurologically identical parallel-world copies of 
ourselves.  This is possibly not an untenable view (so long as we hold on to the 
idea of worlds as fundamental) but to me it seems an unattractive one. 
(The only way to avoid this problem would be to take the preferred basis
as a decoherence basis (see section \ref{persist}) for the brain, in
which case interference between the different terms will be negligible --- 
but this would require the use of detailed biochemical and
neurophysiological criteria to specify what is supposed to be a basis
which is preferred at the level of basic ontology, and even then the
interference terms will not be completely eliminated.)

This is to say nothing of what lies outside my skull --- but if some object many 
light years away is in a superposition of two preferred-basis states then taking 
the worlds literally implies that there are two sets of parallel worlds with 
the copies of my brain in one set identical to the copies in the other set.  In 
this situation are there two identical versions of me, or should I regard my 
thoughts as supervening on both simultaneously? There will obviously be no 
detectable difference between these two viewpoints, but in the former the state of 
the faraway object will be determinate but unknown and in the latter there will be 
no fact of the matter about its state.

Generally speaking, modern versions of Everett get round this problem as follows:
they abandon
the idea of specifying a preferred basis explicitly, and then recover
it --- in an approximate or pragmatic sense --- from the state and the dynamics by 
one consideration or another.  The usual consideration --- and the one
which will be adopted here --- is decoherence 
theory~\cite{zurek,saundersdecoherence}, which tells us that subsystems of the 
world will have `states' (\ie reduced density operators) diagonalized with respect 
to a certain basis, and that attempts to prepare these subsystems in states not 
belonging to this basis will be virtually impossible.  The basis thus determined, 
however, is \emph{approximate}: many choices of basis will satisfy the decoherence 
property with (for instance) the choice of basis for the spin of a given electron 
being irrelevant unless that electron's spin is entangled with some macroscopic 
system (such as a Stern-Gerlach apparatus).\footnote{Furthermore, it is possible that
it is to some extent anthropocentric: it is impossible for functional
systems like us to prepare the subsystems in states not belonging to the basis,
but it remains unproven that other, wildly different systems would face
the same restriction.  I am grateful to Simon Saunders for this point.}

But in rejecting any idea of worlds as ontologically primary, we have 
returned to the problem at the beginning of this section: if the state 
is fundamental, how are we to understand it?  Hopefully the parallels with 
spacetime are clear, and their implications equally so: we are to understand the 
universal state as
\begin{quote}
an entity with such-and-such mathematical description \textit{which can be thought 
of as a collection of instantaneous worlds together with their Hilbert-space 
amplitudes, no matter that it may be decomposed into such a collection in many 
different ways.}
\end{quote}
As with the instants of spacetime, the point is not that we are directly presented 
with worlds of definite particle position through our senses (we aren't, or not 
exactly), nor that we ourselves live in a single such world (we don't) but that 
we have a conceptual grasp on the idea of such worlds.  We can regard the universal 
state as being made up of worlds and their amplitudes in the same way that we can 
regard spacetime as made up of instants and their metric relations: neither description 
really does justice to the symmetries of the entity being described, but both give 
enough data completely to specify the entity and both give us a conceptual grasp of 
what this entity is.  

In this way, the analogy with spacetime and the many-instants view
allows supporters of Everett to answer the charge that, without some
explicitly preferred basis, their world-view simply does not make sense
(as distinct from the question of whether, for instance, it is
empirically valid).  For the idea of a universal state regarded as
ontologically prior to worlds makes no more and no less sense than 
the idea of spacetime regarded as ontologically prior to instants of time, and
if we insist that the Everett interpretation requires a preferred basis to make
sense then we ought to insist also that relativity requires a preferred
foliation.\footnote{It might be argued that \citeN{barbour99} requires both:
his metaphysics (for reasons connected to spacetime relationism and quantum 
gravity) includes both a preferred foliation and a choice of configuration space 
as the preferred basis.}

\section{Consistent Histories and the Persistence of
Worlds}\label{persist}

Let us examine the concept of this ``effectively preferred basis'' more carefully.  
A basis $\{\ket{i}\}$ can equally be expressed as a partition of unity
\[
\id = \sum_i \op{P}_i
\]
where $\op{P}_i$ projects onto the one-dimensional subspace spanned by \ket{i}.  
We can characterise such partitions as being a collection of projectors which
\begin{enumerate}
\item sum to unity;
\item are mutually orthogonal, $\op{P}_i \op{P}_j = 0$ for $i \neq j$; and
\item project onto one-dimensional subspaces.
\end{enumerate}

We can then understand the idea of an ``approximate basis'' mentioned above by 
dropping the third of these requirements.  Decoherence specifies such a collection 
of projectors, and it is an approximate basis in the sense that any exact basis 
which (regarded as a resolution of unity) is a fine-graining\footnote{If we have 
two resolutions of the identity $\{\op{Q}_i\}$ and $\{\op{P}_j\}$, the latter is a 
fine-graining of the former if each $\op{Q}_i$ is a sum of some of the $\op{P}_j$.} 
of this collection, will consist of states which are stable against the decoherence 
process.  By this, we mean that even if at some time a state (expressed as a density 
operator) is not block-diagonalized by the decoherence projectors, the 
off-block-diagonal matrix elements of the operator will decay on a vanishingly short
timescale compared to the timescale on which the block-diagonal matrix elements 
evolve.  The information encoded in the 
off-diagonal states disappears into entanglement between states of macroscopic 
systems and is effectively inaccessible; 

As is well known, the consistent histories formalism (\citeNP{griffiths,gellmann};
see \citeN{kent} for a recent review)
abstracts this requirement and frees it from direct reference to subsystems and 
to the decoherence process.  A consistent history space is, roughly, a collection 
of coarse-grained bases (with the collection indexed by time, so that there is one 
basis for each moment of time\footnote{Actually, in practice we often discretise 
time rather than using the continuum.}).  The consistency condition ensures that 
we can apply classical probability to the events described by the projectors in 
these bases (in the sense that each history can be assigned a probability, and
if history $A$ is a coarse-graining of histories $B_1, \ldots B_n$ then the 
probability of history $A$ is the sum of the probabilities of its component
histories).

The problem given rise to by this abstraction is that there exist many choices of 
consistent history space, but if we follow Everett and keep the state as 
fundamental there is no problem.  Just as our choice of world-decomposition (\ie 
fine-grained basis) is made for ease of description rather than more fundamental 
reasons, so our choice of history space is just made so as to give a convenient 
description of the quantum universe.  In fact there will be a subset of history spaces which 
are \emph{much} more convenient: we are information-processing systems, and it can 
be shown that any such system picks out a consistent history 
space~\cite{saundersevolution}.  (Reverting to the subsystem description given 
earlier, the point is (in part) that such a system needs to store memories and if it 
chooses an encoding of memories into states which are not diagonal in the 
decoherence basis, they will not last long~\cite{halliwell,zurek}.)  So for describing 
events in our vicinity, at least, there is an overwhelmingly preferred choice.  
As with the pragmatically preferred reference frames of relativistic spacetimes, 
the preference is only approximate and really only extends to our spatial vicinity: 
if we wish to pick a truly global, fine-grained basis then there will be 
considerable arbitrariness.

Let us now reconsider the  four problems with many-worlds identified in the 
previous section.  Of these, three have close analogues in the many-instants 
description of spacetime: in describing reality there is neither an
uniquely best set of instants, nor of worlds; 
we cannot recover spacetime from its instants without their temporal 
relations
any more than we can recover the universal state from the worlds without their
amplitudes; and mental facts supervene neither on facts about a single instant
nor about a single world.  Hence if we accept the many-instants description of
spacetime regardless, none of these problems prevent us taking the Everett 
interpretation seriously, nor explicating it (with appropriate care) in terms of
worlds.

However, the third problem identified in section \ref{qm} remains: the 
(fine-grained) worlds in a superposition are instantaneous, and we have not yet
addressed the question of how (or whether) we can track a given world from
moment to moment.  But this situation should be familiar to 
us from spacetime physics: there, too, the notion of a persisting object is not 
directly present in the formalism. However, the notion can emerge given a 
sufficiently ordered spacetime: we may be able to pick out `world-tubes' of fairly 
stable matter configurations, and declare different three-dimensional slices of 
that tube to be the same object at different times.  This concept may not be 
definable with arbitrary precision (how much is an object allowed to change before 
ceasing to count as \emph{the same} object?) and will not be applicable if the 
spacetime is not ordered enough for persistent world-tubes to exist; nonetheless 
it is useful.

We can use the same pragmatic concept of continuance to make sense (in certain 
circumstances) of the idea of identifying worlds across time.  In neutron 
interferometry, for instance, we have a neutron in a linear superposition of two 
spatially separated wave-packets.  Until they are brought back together again 
these packets do not overlap, so the evolution of each packet in
position space can be treated independently, without allowing for
interference between the two, and so we can reasonably describe the 
neutron as being in a superposition of two histories, each fairly well-defined.  This 
makes it reasonable to speak of two persisting worlds, each describing one neutron. 
(See \citeN{vaidman} for a detailed discussion of neutron
interferometry in the Everett interpretation, from a related viewpoint.)
Of course it is vital to remember that this language does not describe anything 
fundamental in the theory, and that it will fail in certain circumstances 
(in this example, it will fail when the two neutron beams are brought back 
together again, so that interference occurs between the two  worlds) or when 
looked at too closely (in the example, we may speak of two persisting worlds, 
but the worlds are fairly coarse-grained).  Nonetheless the concept of persisting 
worlds may be useful in certain explanatory contexts (rather like the concept of 
`living creature' in biology, or `star' in astrophysics, neither of which have an 
unambiguous definition or are written directly into the 
formalism\footnote{See \citeN{deutschfabric} and \citeN{dennettpatterns}
 for further discussion of this issue.}). 

In this context we see the consistent-histories formalism from another viewpoint: 
consistency gives a criterion for (coarse-grained) worlds for which there exists 
a fairly robust notion of persistence.  By `fairly robust' I mean that although we 
may need to speak of worlds 
splitting\footnote{The idea of worlds splitting will obviously lead to ideas of a 
one-many criterion for identity over time, both for the worlds and for the 
macroscopic objects within them.  Though there are objections to such criteria, 
especially
as regards the interpretation of probability (see, \egc, \citeN{barrett}), they have
been discussed in detail elsewhere (in particular by \citeN{saundersprobability} and 
\citeN{tappenden1999}) and will not be 
dealt with further here.} and (in thermodynamically implausible 
circumstances) recombining, we do not have to deal with the interference 
between worlds that generally makes persistence meaningless.

\section{The Status of the Observer}\label{observe}

As was stated in the previous section, if we have a consistent-history basis (at 
 time $t$) $\{\op{Q}_i(t)\}$ then it is constrained by the 
requirement that our brains (being information-processors) carry out this 
processing in such a 
basis.   Then I can identify a set of projectors $\{\op{R}_j\}$ which are a 
coarse-graining of $\{\op{Q}_i(t)\}$ and which project onto functionally 
distinct states of my brain.\footnote{If we are interested in questions of 
supervenience, it may well be that mental facts are supervenient upon 
facts about processes, not states.  In this case we would need to consider 
sequences of coarse-grainings.}  This set of projectors is not a resolution 
of the identity, since that would be to assert that my brain exists with 
probability one; furthermore it is \emph{very} coarse-grained in comparison 
with $\{\op{Q}_i(t)\}$.  After all, specifying a set of projectors onto 
states of my brain does not involve giving any information about the rest of 
the universe.

 We now have a three-level description of the quantum state.

At the lowest level, we have a description in terms of \textbf{Everett 
worlds}, \ie by means of a fine-grained basis.  It is at this level that we 
are giving a \emph{complete} description of the state; nonetheless these 
worlds are instantaneous entities with no traceable histories, and there is 
a high degree of arbitrariness about the choice of basis.\footnote{As 
\citeN{mermin} has shown, we can equally well give a description at this 
level in terms of local events and the correlations between them: given 
any resolution of a system into subsystems, knowing the correlations 
between observables in the subsystems gives us sufficient information 
to recover the state.}  

The arbitrariness is not complete, however, since to be practically useful 
this basis must be a fine-graining of some \textbf{decoherence basis}, 
which will in turn be given by some choice of consistent history space.  
At this level of the description we can talk usefully of histories, 
describing a branching (and occasionally recombining, in principle but not 
in practice) set of worlds.  This is the level at which we obtain a useful 
classical limit.  However a description of the world at this level is not 
complete, since the coarse-grained nature of the decoherence basis means 
that information is lost.

In turn, the choice of consistent history space is motivated by anthropic 
considerations: we are information-processing systems and as such must be 
embedded in some consistent history space in which this information 
processing takes place \cite{halliwell,saundersevolution}.  
As such a useful choice of consistent history space must be a fine-graining 
(at each moment of time, or for each short space of time if we wish to think 
of thoughts supervening on processes rather than states) of the 
\textbf{basis of functionally distinct brain states}.  At this level our 
description of the state is highly incomplete.   However, at least from an 
epistemological view point if the universal state is \ket{\psi} and I am in 
brain state $j$ (with $\op{R}_j$ being the projector onto this state) then 
I should regard the state of the Universe relative to me as given (up to 
normalisation) by $\op{R}_j \ket{\psi}$. 

Note the contrast between the way in which observers are recovered 
from unitary quantum mechanics in
this sort of approach (and in the closely related approaches of, 
\egc, \citeN{saunderslockwood}, \citeN{zurek}, \citeN{gellmann}), compared to the approach taken 
by many-minds theorists like Lockwood \citeyear{lockwoodbook,lockwoodbjps1} or Donald
\citeyear{donaldfoundations,donald,donaldrecent}.\footnote{\citeN{albertloewer}, though they also advocate a 
many-minds theory, differ from Lockwood and Donald in that they add explicit 
extra structure in the form of continuing minds.}
In the latter approach, the projectors $\op{R}_j$ are introduced from the 
outset, and the interpretation  is constructed in terms of them; 
in the former, the job of physics is taken to be the recovery of a 
quasiclassical domain and the problem of consciousness is then handed over 
to other disciplines.

To some extent, reasons for pursuing one or other scheme come down to 
differences of opinion about how successful certain scientific or 
philosophical programs will be, so that decoherence-based approaches depend 
on the success of the decoherence program to succeed formally in recovering 
an approximately classical physics, while many-minds theorists can ignore 
this program but are instead committed to certain quite strong physical or 
philosophical views about consciousness (so Lockwood's approach requires 
there to be a fact of the matter as to which physical systems possess 
consciousness bases,\footnote{cf.\,\citeN{lockwoodbjps2}:
\begin{quote}
I regard 
it as a straightforward matter of fact whether cats or (a more seriously 
contentious issue) prawns are sentient.  Thus it's not, as I see it, a 
matter of how we define our terms, but whether there's a `what it's like 
to be' a cat or a prawn [\ldots]
\end{quote}   If, in fact, the correct theory of 
consciousness were to be a much less all-or-nothing affair (\eg 
\citeN{dennettkinds}:
\begin{quote}
the features [of animal consciousness] we have examined seem 
to make their appearance not just gradually but in an unsynchronised, 
inconsistent and patchy fashion, both in evolutionary history and in the 
development of individual organisms.)
\end{quote} it would presumably have a 
significant impact on Lockwood's theory.} while 
\citeN{donaldfoundations} has constructed explicit models of observers 
and holds that this is a necessary part of interpreting 
Everett\footnote{cf.\, \citeN{donald}:
\begin{quote}
[\ldots] the proper task of an 
analysis of quantum theory is to give an exact definition of the possible 
physical manifestations of an observer.\end{quote}}).

It is obviously far too early to say which way these longstanding 
controversies will resolve themselves; however, section \ref{theories} 
seems to give some support to the decoherence approach: whether or not 
mind turns out to be essential in interpreting quantum mechanics, we 
will have to recover a quasiclassical domain \emph{anyway} to make 
contact with existing theories and so interpret quantum theory as 
a theory of physics.  If there are no theory-neutral observations, then 
saving the observed phenomena requires us to save (something approximately 
isomorphic to) the theory.

I mention in passing an ambiguity about our choice of $\op{R}_j$.  
These projectors are designed to pick out \emph{my} brain, but how is that 
to be done?  Presumably my thoughts can be instantiated in many different 
physical media; they can certainly exist in many different spatial locations 
and be composed of different atoms, so requiring them to project onto a 
given subsystem's Hilbert space isn't really plausible.  Presumably I have 
to use some intrinsic structural features, but there seems to be an 
arbitrariness as to how great a personality-shift can be tolerated before 
a given projector should actually be counted as projecting onto 
\emph{someone else's} brain state.  This is, of course, the old problem 
of personal identity in quantum-mechanical guise.

\section{Covariant Quantum Mechanics}\label{covqm}

So far, we have constructed an analogy between spacetime, and the
quantum state \emph{at a given time}.  However, talking of worlds at a
given time obviously implies a choice of reference frame and a breaking of 
covariance, whereas  one of the major motivations for adopting the Everett approach 
is its claim to 
provide a quantum mechanics which is compatible with relativity.
Accordingly, we now turn to the problem of understanding relativistic
quantum theory.

The mathematical structure given for quantum field theory at the beginning of 
section \ref{qm} is fully covariant, since it is written in the 
Heisenberg formalism so that its states are not dependent on time.  Hence we have 
an uninterpreted object --- the state \denop\ --- which has the properties we 
require.  To describe this object we can combine the many-worlds and many-instants 
descriptions:
\begin{quote} 
The universal state is an object with such-and-such mathematical description 
\textit{which can be described as a collection of sets of worlds, together with 
the temporal metric relations between sets and the amplitudes of individual worlds 
within each set, no matter that the universal state can be decomposed into such a 
collection in many different ways}.
\end{quote}
There is an attempt to illustrate this viewpoint on page 277 of 
\citeN{deutschfabric}, although this illustration gives the impression that 
individual worlds have histories (\ie that it is possible to say that \emph{this} 
world in \emph{this} instantaneous collection of worlds is the future version of 
\emph{that} world in \emph{that} collection).  This is consistent with Deutsch's 
explicit introduction of a preferred basis in \citeN{deutsch85}, and with the 
`continuing minds' of \citeN{albertloewer} (in both of which there exist
persisting objects --- be they minds or worlds --- which are stochastically
partitioned as the wave-function evolves), but contrasts with the view presented 
here, in which persisting worlds are an approximate and derived notion.  
(Obviously we could justify such a description in this framework by working at the 
level of 
consistent histories, but then we cease to have a complete description of the full 
instantaneous collection.)  Deutsch also makes the speculative suggestion 
(inspired by canonical quantum gravity and by the work of \citeN{pagewootters}; 
see also \citeN{barbour99}) that this two-axis collection of worlds can be 
collapsed into a single collection --- in other words, that different instants 
can be understood as different Everett worlds --- but this remains controversial, 
and lies beyond the scope of this paper.

It must be admitted that our description of the universal state, 
given as it is in terms of global instants of time
and worlds defined at those instants, is somewhat unaesthetic.  In the next two
sections I shall develop a somewhat modified version which better reflects relativistic
covariance, and then consider an important strength of our existing description.

\section{Spacetime as the Set of Events}\label{events}

One partial method of emancipating special-relativistic physics from the language 
of space-at-an-instant is to see it as the set of events and 
the spatiotemporal relations between them.\footnote{Specifying these spatiotemporal 
relations is crucial: without them we have no way of distinguishing numerically 
distinct, but identical, events (such as two identical flashes of light in 
different places).  In this (rather Leibnizian) sense an event is really a global 
notion --- to specify it, we need to specify its relations with all other events.}
This approach deserves mention here, as it succeeds partially in conveying a direct 
intuition of relativistic spacetime but is still reliant in many ways on our 
prerelativistic intuitions.  

The easiest way to see this is to look more closely at the concept of `event' --- 
by which we mean something that occurs in a given place, at a given time 
(an explosion, say, or a flash of light, or the coincidence of hands on a clock).  
But this description already makes intensive use of prerelativistic concepts, both 
in its direct use of space and time language and in its reference to physical 
objects.  We could of course define events as crossing-points of world-lines, but 
now we have either to explain what world-lines represent (surely impossible without 
our prerelativistic language) or to return to uninterpreted mathematics 
(`a world-line is a smooth map of the real line into the four-manifold such that 
the metric distance of any two points in the image is positive', 
etc.\,).\footnote{The subtleties of defining spatially and temporally extended events were 
recognised by \citeN{einstein05}, whose original discussion of special relativity 
refers in passing to
\begin{quote}
the imprecision that is inherent in the concept of simultaneity of two events 
taking place at (approximately) the same location and that must be surmounted by 
an abstraction.
\end{quote}  
The issue is discussed further by \citeN{saundersmetaphysics}.
}

However, one thing this discussion reminds us of is that our use of prerelativistic 
intuitions does not require a \emph{global} notion of space.  We cannot say what 
we mean by an event without any reference to the space/time split but we can 
describe it with reference to space and time only in its vicinity.  

As such we may understand relativistic spacetime by 
\begin{enumerate}
\item breaking it up into small regions;
\item describing each region in terms of a (highly non-unique) space-time split 
--- \ie a local version of many-instants;
\item giving the spatiotemporal relations between the regions.
\end{enumerate}
The pure-event viewpoint can be understood as a limiting case of this 
interpretation scheme, although in the limit we once again lose our ability 
to understand the theory (as has been argued for in section \ref{spacetime}).

This method of understanding relativistic spacetime has a number of advantages over 
the use of 
globally defined instants: in particular it brings out the arbitrariness of our 
choice of spatial slice at large distances and the irrelevance of that choice in 
describing local phenomena, and it makes it clear that there is no fact of the 
matter about what is happening \emph{now} in distant regions of space.  However, 
it is not really a more ``right'' description of spacetime --- the best which we 
might be able to say about it is that it is a more useful description of an entity 
whose perfect description as a physical system lies (at least for the moment) 
beyond our ability to comprehend directly.

The set-of-events description of spacetime
can be copied in the quantum case.  An (instantaneous) `event' in 
quantum theory is specified by a (Heisenberg-picture) projector (onto states 
for which, at the time at which the event is supposed to occur, it does indeed 
occur with certainty).  If events $A,B$ are described by Heisenberg projectors 
$\op{P}_A, \op{P}_B$ respectively, then the probability of event $A$ occurring 
given that event $B$ occurs is 
\[
\frac{\tr(\op{P}_A \op{P}_B \denop \op{P}_B \op{P}_A)}
{\tr(\op{P}_B \denop \op{P}_B)}.
\]
We are now in a position to regard the quantum state as (being described by) 
the collection of events together with the spatiotemporal and probabilistic 
relationships between them.  

This relational view of Everett has been recently championed by 
Saunders~\citeyear{saundersevolution,saundersdecoherence,saunderstense,saundersmetaphysics,saundersprobability}.  As with the classical description of 
spacetime as events and spatiotemporal relations between them, it is in many ways 
less arbitrary than using global `instants' and `worlds' but still relies on a 
(classical, pre-relativistic) concept of what an event is.  Essentially the 
concept which we import from our previous theories is that of a `local world', 
\ie a small spatial region, at a single time, in which certain observables have 
definite values (note the explicit need to choose a local simultaneity convention,
for self-adjoint projectors can project onto spatially extended events but not
onto temporally extended ones).\footnote{Recent work by 
\citeN{deutschhayden} on the Heisenberg picture in quantum computation has given
rise to some suggestions of another entirely local --- and resolutely non-classical ---
way to understand unitary quantum mechanics, but this work remains in its
infancy.}

\section{Telling a Dynamical Story}\label{dyn}

The `local' description given in the previous section would appear to
have significant advantages over the many-instants approach, in that it
better captures the notions of covariance and locality which are central
to relativity.  Should we then abandon the many-instants description of
section \ref{covqm} in favour of the local description?

We can see why such abandonment would be unwise when we consider the issue
of dynamics.  We need the notion of a global 
foliation of spacetime (whether we are doing classical or quantum physics) 
if we wish to discuss spacetime physics as a dynamical 
system, and in particular if we wish to discuss issues of determinism.  A globally 
specified instant, together with the (classical or Schr\"{o}dinger) state at that 
instant, incorporates an important feature of our prerelativistic theories 
which is less transparent in the local description: namely that at any given moment 
of time we can specify the contents of space everywhere without having to worry 
that one region has causal influence upon another, but that having done so, at 
later moments of time the contents of space will be determined by the dynamics.

In discussing dynamics in relativity, we are again required to behave in this way: 
we need to specify the matter distribution and its rate of change across a 
spacelike surface (\iec a choice of instant) which extends across the whole region of 
interest (in principle across the whole universe) and then evolve this spacelike 
surface forward in time~\cite{wald}.  There are, of course, a huge number of 
different ways in which we may specify this time evolution, each generating a 
different choice of spacetime foliation and hence a different `many-instants' 
description of spacetime (this is the `many-fingered time' of \citeN{MTW}). 

The covariance of relativity theory manifests itself in our ability to describe a 
dynamical process equally validly by many different foliations.  However, only 
in the most stylised of dynamical processes in classical or quantum 
relativity can we 
give such a description in a manifestly covariant way which makes no use of a 
foliation\footnote{Contrast the two-body collisions
of otherwise free particles (adequately described in terms of world-lines) with
computer models of supernovae \cite{arnett}, or the (admittedly complex) scattering theory of
asymptotically free relativistic quantum particles with the far more intractable 
lattice simulations used to analyse non-perturbative quantum field theory (see
\citeN{kaku} for an introduction).}: the very concepts of `process' and `dynamics' depend upon the notion of 
something taking place as time changes, and so cannot be given without reference to 
time.

As such, if we wish to describe causal or dynamical processes in a given region, we 
must make a space/time split which is global at least over that region.  So if we 
were describing the evolution of life on earth, for instance, we may be able to 
restrict our slices to some neighbourhood of the solar system --- and indeed can 
get quite far by restricting them to a neighbourhood of the Earth's surface.  Only 
when we do cosmology do we need to foliate the entire spacetime.

How much freedom do we have in choosing our foliation, and our set of worlds?
For a large class of dynamical processes (essentially all those taking place in
our local vicinity, at macroscopic scales and low velocities) there is (locally) 
a fairly well defined `best choice' of foliation and of basis common to all
members of the class: the foliation picked out by Lorentz reference frames 
co-moving with the Earth, and the basis picked out by decoherence.  Many other
processes, while not sharing these choices, still have an obviously `best'
choice: for classical relativity, consider the atmospheric physics of a star
moving at relativistic speeds relative to us; in quantum physics, consider
neutron interferometry with its fairly well-defined separate histories.  Other
processes again cannot be properly understood until we have seen them described
in several distinct bases or foliations: consider:
\begin{itemize}
\item colliding shock-waves, where we need to consider both the centre-of-mass
viewpoint and that of an observer moving with the shock;
\item gravitational collapse \cite{MTW}, where both the description from a 
surface co-moving with the collapsing star and the description of a faraway 
observer give
important insights;
\item the two-slit experiment, which notoriously cannot 
fully be understood from either a particle or a wave viewpoint;
\item certain algorithms in quantum computation \cite{deutschekert}
(indeed, from a certain viewpoint every algorithm in quantum computation, since
quantum computers outpace classical computers partly through not being forced to
work in a fixed basis).
\end{itemize}
The moral is that we should not regard `local' and `many-instants'
descriptions, whether of classical or quantum physics, as rivals:
rather, they illuminate different aspects of the (classical or quantum)
universe, and which one is used will depend on the specific situation.

\section{Summary: a Point-by-Point
Comparison}\label{table}

The rich analogies between spacetime and the Everett interpretation stand out
clearly in the following table:

\begin{center}

\begin{tabular}{|p{60mm}|p{60mm}|}
\hline Relativity & Everett \\ \hline \hline

The Universe may be fully described as a collection of instants and
their temporal relations. &
The Universe may be fully described as a collection of (fine-grained) worlds, and their weights. 
 \\
& \\
Such a description is arbitrary and fails to show the full structure of spacetime.  & 
Such a description is arbitrary and fails to show the full structure of the multiverse.\\
\hline
The instants are theoretical constructs and our thoughts do not supervene on individual instants. & 
The worlds are theoretical constructs and our thoughts do not supervene on individual worlds.  \\
&  \\
Nonetheless our familiarity with our pre-relativistic theories gives us a conceptual
grasp on the idea of `now' and an instant,which enables us to understand what is described by spacetime. & 
Nonetheless our familiarity with our pre-quantum theories gives us a conceptual 
grasp on the idea of a world which enables us to understand what is described by quantum theory. \\
\hline
Specifying only the collection of instants without giving the temporal relations between instants is 
insufficient to tell us the spacetime.\footnotemark &
Specifying only the collection of worlds without giving the amplitudes of each world is insufficient 
to tell us the state. \\
&\\
We know mathematically how to handle spatiotemporal relations, and we have a practical understanding 
of temporal duration and flow, but these concepts are philosophically problematic. &
We know mathematically how to handle amplitudes, and we have a practical understanding of probability, 
but these concepts are philosophically problematic.\\
&\\
We may speak of `moments of time' and the number of moments of time (`the next moment', \etc) 
but this is just a metaphor for temporal duration, 
and cannot be interpreted literally.&
We may speak of `number of worlds' but this is just a metaphor for the weight of a given world, 
and cannot be interpreted literally. \\ 
\hline

\end{tabular}
\end{center}
\footnotetext{See Barbour and Bertotti~\citeyear{barbourbertotti77,barbourbertotti82} and \citeN{barbour99}, however, for further discussion of this point from a Machian perspective.}

\begin{center}

\begin{tabular}{|p{60mm}|p{60mm}|}
\hline Relativity & Everett\\ \hline \hline

The theory is specified in a way which makes no direct reference to observers, 
but to understand why the Universe seems as it does to \emph{us} we need to address certain observer-related problems 
--- specifically we need to understand our perceptions of the passage of time.\footnotemark&
The theory is specified in a way which makes no direct reference to observers, but 
to understand why the Universe seems as it does to \emph{us} we need to address certain observer-related 
problems --- specifically we need to understand our perceptions of probability. \\
\hline
In everyday circumstances, there exists an (approximate) natural choice of
choice of spacetime foliation.\addtocounter{footnote}{-1}\footnotemark &
In everyday circumstances, there exists an (approximate) natural choice of basis.
\\
& \\
The details of this foliation are arbitrary, both when examined  closely and with respect to spatially 
remote areas.\addtocounter{footnote}{-1}\footnotemark&
The details of this basis are arbitrary, both when examined closely and with respect to spatially remote areas. \\
\hline
In describing the dynamics of a system it is generally necessary for us to give our 
description in terms of some choice of foliation (\iec instants).&
In describing the dynamics of a system it is generally necessary for us to give our description in 
terms of some choice of basis (\iec worlds).\\
&\\
For some processes (such as a boardroom meeting or the dynamics of the Solar 
System) there is an approximately-defined `best' choice of foliation; for others 
(such as gravitational collapse) different choices may give different insights into the process.&
For some processes (such as neutron interferometry, or a Schr\"{o}dinger's Cat 
experiment, or (possibly) making sense of counterfactual reasoning) there is an
approximately defined 
`best' choice of worlds; for others (such as a quantum computation) different choices 
may give different insights into the process. \\
\hline

There is no fundamental notion of an object's persistence through time; 
all we have are structural similarities between regions of space at different times.\footnotemark &
There is no fundamental notion of a world's persistence through time; 
all we have are structural similarities between parts of the state at different times.\\    
&\\
However, in certain situations it may be possible to recover a (pragmatic, approximately defined) notion of 
persistence of objects from these structural features.\addtocounter{footnote}{-1}\footnotemark &
However, in certain situations it may be possible to recover a (pragmatic, approximately defined) notion of 
persistence of worlds from these structural features.\\
\hline

\end{tabular}
\end{center}
\addtocounter{footnote}{-1}
\footnotetext{These matters are discussed further by \citeN{stein91}.}
\addtocounter{footnote}{1}
\footnotetext{It seems likely that these notions apply to our own concept of personal identity as much as to physical objects; see \citeN{parfit84} for a defence of this.}
\begin{center}

\begin{tabular}{|p{60mm}|p{60mm}|}
\hline Relativity & Everett  \\ \hline \hline

We may describe the classical spacetime as a network of events together with the spatiotemporal 
relations between the events.& 
We may describe the quantum spacetime as a network of events together with the spatiotemporal and 
probabilistic relations between the events. \\
& \\
Whether we describe spacetime in terms of many global instants or in terms of more localised events, 
we make essential use of our intuitive concept of a local spatial region --- not in the mathematical 
formulation of the theory but in getting a conceptual grasp of it and understanding our own place in it. 
&
Whether we describe the quantum universe in terms of many global worlds or in terms of more localised events, 
we make essential use of our intuitive concept of a local and value-definite (for some observables) spatial 
region --- not in the mathematical formulation of the theory but in getting a conceptual grasp of it and 
understanding our own place in it.
\\
\hline
\end{tabular}
\end{center}
\section{Conclusion}\label{concl}

Are the worlds real?  Yes, in the sense that instants of time are real: they 
may not be present directly in the formalism, but unless we introduce the 
concept we may struggle to understand what the formalism is telling us, and 
without the concept it may be impossible to capture important 
(causal/deterministic) properties of the world.  Further, a description in 
terms of worlds (or instants of time) is not \emph{incomplete}, for we can 
certainly recover the universal state from it; our only complaint with it is 
that it is somewhat arbitrary.

Are consistent histories, and worlds which persist over time, real?  Yes, 
in the sense that rivers, or animals, or persisting objects, are real: like 
worlds or instants they are not directly present in the formalism, and 
unlike worlds or instants they only approximately definable, but that is 
no reason why they should not be seen as legitimate entities or used in 
our explanations (any more than we should expect to be able to describe 
zoology in any useful or explanatory way using only the language
 of quantum field theory).

We are undoubtedly more at home with Minkowski spacetime than with the 
universal state.  Partly this may be because we have worked with the concept 
in physics for rather longer, but more importantly we have long been used 
to the idea that multiple times exist (in some sense) --- the innovation in 
relativity theory is the unification of these instants into a whole, and 
the identification of the instants as secondary concepts.  Everett asks us 
to take both steps at once: to accept that there exist many 
worlds,\footnote{Of course, this is not a wholly new concept: whatever we 
may think of the metaphysical status of possible worlds, we routinely 
describe them when we use counterfactuals.} and then to fuse them together 
into a whole and accept that the worlds are only secondary.  Clearly this 
is a significantly larger conceptual jump; still, if we are prepared to 
accept the existence of many worlds and if we are happy with the step from 
many times to spacetime, there seems no reason to avoid a similar step in 
the case of quantum theory.

\small
\vspace{0.5cm}

\noindent \emph{Acknowledgements}---For extensive discussions, suggestions and comments on this work, I am very 
grateful to Hannah Barlow, Katherine Brading, Harvey Brown, Jeremy Butterfield, Chris Fuchs, Clare 
Horsman, Lev Vaidman, and especially to Simon Saunders, whose work on Everett has been a
major influence on this paper.  I am also grateful to David Deutsch for many discussions of the Everett 
interpretation.

\end{document}